\documentclass[fleqn,usenatbib]{mnras}

\usepackage{float}
\usepackage{amssymb, amsmath, epsfig}
\usepackage{graphics,times,array,ctable,amsmath,amssymb,natbib}
\usepackage{threeparttable}
\usepackage[export]{adjustbox}
\def\apj{ApJ}
\def\apjl{ApJL}

\def\aj{AJ}

\def\mnras{MNRAS}

\def\araa{{Ann.\ Rev.\ Astron.\& Astrophys.\ }}
\def\aap{{A\&A}}

\title[UVB Tomography with CASTOR and SPHEREx]{Forecasts for Broadband Intensity Mapping of the Ultraviolet-Optical Background with  CASTOR and SPHEREx}

\author[Scott, Upton Sanderbeck, and Bird]{Bryan R Scott,$^{1}$\thanks{bryan.scott@email.ucr.edu}
Phoebe Upton Sanderbeck,$^{1,2}$
Simeon Bird$^{1}$
\\
$^1$Department of Physics \& Astronomy, University of California, Riverside, 900 University Ave, Riverside, CA 92521, USA\\
$^2$Los Alamos National Laboratory, Los Alamos, NM 87545, USA}

\begin{document}
\label{firstpage}
\pagerange{\pageref{firstpage}--\pageref{lastpage}}
\maketitle

\begin{abstract}
Broadband tomography statistically extracts the redshift distribution of frequency dependent emission from the cross correlation of intensity maps with a reference catalog of galaxy tracers. We make forecasts for the performance of future all-sky UV experiments doing broadband tomography. We consider the Cosmological Advanced Survey Telescope for Optical-UV Research (CASTOR) and the Spectro-Photometer for the History of the Universe, Epoch of Reionization, and Ices Explorer (SPHEREx). The dominant uncertainty is from variability in the photometric zero point, which scales with limiting magnitude and so mirror size. With this scaling, and assuming a galaxy number density characteristic of future spectroscopic datasets, we find that CASTOR measures the UV background SED $2-10$ times better than existing data. The applicable redshift range will expand from the current $z < 1$ to $z \approx 0-3$ with CASTOR and $z =5-9$ with SPHEREx. We show that CASTOR can provide competitive constraints on the EBL monopole to those available from galaxy number counts and direct measurement techniques. At high redshift especially, these results will help understand galaxy formation and reionization. Our modelling code and chains are publicly available.
\end{abstract}


\section{Introduction} \label{sec:intro}

The extragalactic background light (EBL) is a powerful probe of structure formation, cosmic star formation history, and the intergalactic medium \citep{Overduin_2004, McQuinn_2013}. Because the ultraviolet background (UVB) component of the EBL is both a direct probe of these processes and sets the ionization state of the intergalactic medium, understanding the evolution of the UVB and the EBL is both an important modeling problem and a promising observational constraint on the photon distribution. Contributions to the EBL (and UVB) come from direct emission due to galaxies and active galactic nuclei that produce a discrete component, and from radiative processes, including dust scattering and recombination that produce a diffuse component. The EBL contains information about the emission over cosmic time of total integrated processes, and therefore provides an important consistency check for models that attempt to reproduce the photon production history of the universe \citep{Hauser}.

In the ultraviolet (UV), direct measurements of the EBL have been attempted with Voyager 1 and 2 at 110 nm \citep{Murthy1999}, and with Voyager UVS at 100 nm \citep{Edelstein2000}. At these wavelengths, the dominant foreground uncertainty is due to skyglow, leading to large uncertainties on the total EBL intensity \citep{Mattila2019}. In the blue portion of the optical, attempts have also been made with Pioneer 10 and 11 \citep{Matsuoka2011} at 440 nm with the broadband Long Range Reconaissance Imager (LORRI) on New Horizons from 440 nm - 870 nm \citep{Zemcov2017}. Direct photometric measurements with New Horizons have found that a large fraction of background photons may be associated with a diffuse component not associated with identified sources \citep{lauer2020new}. This diffuse component may arise either from galactic sources or extragalactic sources below the limiting magnitude of current galactic surveys. 

Beyond spacecraft measurements, observations as a function of galactic latitude \citep{Hamden2013,Murthy2018} place upper limits on the total combined galactic and extragalactic component, subject to foreground uncertainties due to the zodiacal component and contributions from faint stars. One way to mitigate these uncertainties is the dark cloud technique \citep{Mattila1990,Mattila2017}, where observations are taken in the direction of opaque clouds in the interstellar medium and compared to blank sky observations. The difference between these two measurements is taken to estimate the foreground and combined brightness. Each of these techniques attempts to capture the contribution of diffuse emission at low surface brightness, which results in large statistical and systematic uncertainties arising from the larger skyglow, zodiacal, and galactic foregrounds. The latter also complicates interpretation of direct intensity constraints as they must be decomposed into galactic and extra-galactic components.

An alternative strategy (for a recent review see \citealt{Hill2018}) integrates the total emission over number and luminosity counts of sources down to some limiting magnitude \citep{Madau2000,Gardner2000,Driver2016}. This derives an indirect lower limit on the extragalactic component, subject to large uncertainties in the contributions of fainter and diffuse sources. Since constraints on the photon production history of the universe require knowledge of the diffuse component that is produced by both undetected sources and scattering far from detected discrete sources, significant uncertainties remain about the relative contributions.

To measure the total contribution of faint sources, intensity mapping measures a continuous spatial brightness function on the sky without setting an absolute detection threshold. Using the intensity maps in the Galaxy Evolution Explorer ({\small GALEX}) near (NUV) and far ultraviolet (FUV) from \citet{Murthy2010} and \citet{Murthy2018}, \cite{Chiang_2019} introduced the concept of broadband intensity tomography to measure the EBL.  By cross correlating a spectroscopic tracer catalog with these maps \citep{Newman_2008,mnard2013clusteringbased}, an integrated constraint on the total EBL in two filters was derived with high fidelity separation between the extragalactic signal and the foreground and galactic contributions. In this work, we forecast a similar measurement using future UV telescopes. We consider in particular the Cosmological Advanced Survey Telescope for Optical and UV Research ({\small CASTOR}), a one meter class telescope intended for launch in the mid to late 2020s \citep{Cote2019}, and Spectro-Photometer for the History of the Universe, Epoch of Reionization and Ices Explorer ({\small SPHEREx}).  {\small CASTOR} is a wide field of view survey satellite capable of producing all sky intensity maps with an expected $0.15$'' PSF. {\small SPHEREx} is an infrared observatory which can extend and complement local restframe UV measurements by observing at higher redshifts, $z \approx 5-9$.

The nominal mission design for {\small CASTOR} includes a survey over a $7700$ deg$^2$ \citep{Cote2019} region defined to cover the overlap between the Roman Space Telescope \citep{Dore2018}, Euclid \citep{laureijs2011euclid}, and Vera C. Rubin Observatory \citep{ivezic2019} survey areas. It will provide complementary information in the optical and near ultraviolet to those surveys targeting longer wavelengths. The CASTOR surveys would be performed in three broadband filters from $150$ nm to $550$ nm. The larger mirror size, overall redder filter set and improved calibration compared to the $<250$ nm NUV and $<150$ nm FUV filters on {\small GALEX} offers potential for extending integrated  constraints  on  the  UV-optical background light,  including Lyman continuum (LyC) escape fractions, the slope and normalization of the continuum emission and its evolution to intermediate redshifts up to $z \approx 3-4$, with improved error properties at lower redshift.

{\small SPHEREx} is an all sky spectro-photometric survey covering 0.75-5 $\mu$m. The spacecraft features a 0.2 m mirror and produces spectra of 6.2 arcsec$^2$ pixels by scanning 96 linear variable filters across the sky. Although nominally an infrared survey, at $z>5-9$, {\small SPHEREx} will produce rest-UV intensity maps of Ly$\alpha$ emission. Modeling the complete response of a set of narrowband filters approximating the {\small SPHEREx} instrument can extend these maps to measure the UV continuum and its evolution across these redshifts.

The plan of this paper is as follows. In Section~\ref{sec:1}, we introduce our notation and derive the cross-correlation function which is at the heart of broadband tomography. In Section~\ref{sec:RT}, we introduce our emissivity model and the application of this technique to the expected CASTOR throughput and wavelength coverage.  Section~\ref{sec:error} estimates the error budget for such a survey.  In Section~\ref{sec:results}, we discuss how the additional filter and different wavelength coverage impact our inference of the underlying spectral energy distribution and the EBL monopole.  Finally, in Section~\ref{sec:spherex}, we extend these results with a discussion of how a broadband tomographic measurement with CASTOR is complementary to broadband constraints from {\small SPHEREx} and {\small LUVOIR} on the ultraviolet background and history of Ly$\alpha$ emission. We conclude in Section~\ref{sec:conclusions} with a summary of this work. 

Where necessary, this paper assumes a 2018 Planck fiducial cosmology \citep{Planck2018} with $\Omega_m = 0.31$, $\Omega_{\lambda} = 0.69$, $\Omega_{B} = 0.05$, and  $H_0 = 67$ km/s/Mpc.

\section{Broadband Tomographic Intensity Mapping} \label{sec:1}

In this Section we will introduce our notation and derive the cross-correlation function between photon over-intensity and matter over-density. We will work throughout with quantities averaged over angular shells. Quantities will be a function of $r$, the size of a radial shell on the sky, or $\theta$, the angular distance on the sky. We will also measure change with redshift, $z$, and denote quantities which depend on frequency with a subscripted $\nu$. We follow the derivation from Section 2 of \cite{Chiang_2019}, and summarize below.

\subsection{The Filter Specific UV/Optical Photon Intensity}
\label{sec:jnu}

The comoving emissivity in the restframe, $\epsilon_\nu(r,z)$, often presented in ergs s$^{-1}$Mpc$^{-3}$ Hz$^{-1}$, is not a directly observable quantity, but is related to the observed angle average specific intensity at frequency $\nu$, $j_{\nu}(r,z)$ (in ergs s$^{-1}$ cm$^{-2}$Hz$^{-1}$Sr$^{-1}$) via the equation of cosmological radiative transfer \citep{1993ppc..book.....P};
\begin{align}
\label{cosmo_transfer}
\bigg(\frac{\partial}{\partial \nu} - \nu H(z) \frac{\partial}{\partial t} &+ 3H(z) \bigg) j_\nu(r,z) \\
&= -c\kappa_\nu j_\nu(r,z) + \frac{c}{4\pi} \epsilon_\nu(r,z) (1+z)^3\,.
\end{align} 
Here $H(z)$ is the Hubble function, $c$ is the speed of light and $\kappa_\nu$ is the opacity. The integral solution in the observed frame is 
\begin{equation}
\label{specific_emiss}
j_{\nu, obs}(r) = \frac{c}{4\pi} \int_{0}^{\infty} dz \frac{1}{H(z) (1+z)} \epsilon_\nu(r, z) e^{-\tau_\mathrm{eff}(\nu, z)},
\end{equation}
and $\nu = (1+z)\nu_{\rm obs}$. $\tau_\mathrm{eff}(\nu, z)$ is the effective optical depth, for which we use the simple model of \cite{madau2000intergalactic}, 
\begin{equation}
\label{tau_eff}
\tau_{\rm eff}(\nu, z) = \frac{4}{3}A\sqrt{\pi \sigma_L} \left(\frac{\nu}{\nu_L}\right)^{-1.5} \big((1+z)^{1.5} - 1 \big),
\end{equation}
where $A = 1.4 \times 10^7$ is a normalization constant fixed by estimates of the density of Lyman Limit Systems (LLS), $\sigma_L = 6.3 \times 10^{-18} \mathrm{cm}^2$ scales the hydrogen photoionization cross section, and $\nu_L$ is the Rydberg frequency $3.3\times 10^{15}$ Hz. 

A broadband survey measures the convolution of this quantity with the filter response function $R_Y(\nu_{obs}$),
\begin{equation}
\label{Big_J}
J_Y(r) = \int \frac{d\nu_{obs}}{\nu_{obs}} R_Y(\nu_{obs}) j_{\nu, obs}(r)\,.
\end{equation}
Here $Y$ denotes the filter band and is $u$, $v$ or $g$. $R_Y(\nu_{obs})$ has been normalized such that $J_Y(r) = j_\nu(r)$ for a flat input spectrum. This ensures that the band averaged magnitude is a function of the source emissivity over the observed frequencies, rather than the band shape. 

\subsection{The Broadband Tomography Cross-Correlation Function}
\label{sec:crosscorr}

Intensity mapping experiments measure a continuous spatial flux distribution, rather than a discrete sampling of emitting sources. Thus, rather than constructing a catalog of objects, intensity mapping experiments produce maps of the sky in which the intensity of each pixel is associated with both resolved and unresolved sources.

Broadband tomography \citep{Chiang_2019} is a technique for extracting the redshift distribution of emission from intensity maps in the presence of bright foregrounds. Cross correlating an intensity map with a spectroscopic tracer catalog results in an estimate of the emission distribution in redshift. Here we summarize the key features of this technique and refer the interested reader to previous work for more details. 

Our ultimate aim is to estimate the comoving emissivity as a function of redshift, $\epsilon_{\nu}(z)$, from the band averaged specific intensity, $J_Y$, measured on the sky. We build a model for $\epsilon_\nu(z)$, and forward model  an estimate of the redshift evolution of $J_Y$, denoted $dJ_{Y}/dz$. We compare this forward model to the cross-correlation of the over-intensity observed in the map with a reference catalog and thus estimate the parameters of the model.

To make this explicit, consider a field $X$ with coordinates $(\phi,z)$, where $\phi$ is the angular separation from some reference location $\theta$, and a spectroscopic tracer object. 
If the overdensity of spectroscopic tracer objects is represented by $\delta_r(\phi + \theta, z)$ (which is $1$ if a tracer object is located at a location $\phi + \theta$ and $0$ otherwise), then the cross-correlation is

\begin{align}
\label{eqn: generic_covaraince}
\omega_{X,r}(\theta, z) = \int X(\phi,z) \delta_r(\phi + \theta, z)  d\phi
\end{align}
Subscripts indicate the two fields that are being cross-correlated. Subscript $X$ would be $\epsilon$ for the emissivity map, or $J$ for the observed band-averaged specific absolute over-intensity $\Delta J_Y$. Subscript $r$ indicates the reference emitter catalogue.


To relate the observable cross correlation to the comoving emissivity in redshift, consider the case where we have three dimensional information about the distribution of emitters. Then the comoving emissivity, $\epsilon_{\nu} (\theta,z)$, is given by 
\begin{align}
\label{correlation_function_def}
\epsilon_\nu (\theta, z) &= \epsilon_\nu(z) (1 + \omega_{\epsilon,r}(\theta, z))\,, \\
\epsilon_\nu(z) &= \left<\epsilon_\nu(\theta,z)\right>\,, \nonumber
\end{align}
where $\epsilon_\nu (\theta, z)$ is the specific emissivity at angular separation $\theta$ from objects in the reference catalogue at redshift $z$ and $\epsilon_\nu(z)$ is the spatial mean specific emissivity. Angle brackets denote a 2D average over both sky directions. $\omega_{\epsilon,r}(\theta, z)$ is the angular cross correlation function between emissivity and the spectroscopic tracer catalogue, defined in Equation~\ref{eqn: generic_covaraince}. Equation~\ref{correlation_function_def} is interpreted as giving the excess probability of detecting photons at the given separation. 


Our goal is to relate this to the observable cross correlation between the band-averaged specific intensity map and the spectroscopic catalog, $\omega_{J,r}$. To do this, we substitute the integral solution of Equation~\ref{specific_emiss} into Equation~\ref{correlation_function_def}, and then apply the binning integral from Equation~\ref{Big_J} to find
\begin{align}
\label{angular_correlation_step_1}
J_{Y}(\theta) = \left< J_{Y}(\theta) \right> + \int \omega_{\epsilon,r}(\theta,z) \frac{dJ_{Y}}{dz}(\theta, z) dz\,.
\end{align}

The first term on the right side of Equation~\ref{angular_correlation_step_1} represents the average specific intensity, or monopole, while the second is the term we seek, relating the shell 1-bin averaged map intensity $J_Y$ to the redshift distribution and the rest frame emissivity cross correlation. 
To increase the signal to noise ratio on estimates of the cross correlation, we define normalized weights $W(\theta)$ such that $\int W(\theta) d\theta = 1$ in angular separation and integrate out the angular dependence to produce a one-bin measurement of the cross correlation or another angular dependent quantity.

The second term in Equation \ref{angular_correlation_step_1} can be expressed in terms of the underlying matter correlation function $\omega_{MM}(z)$ if both the intensity map (or comoving emissivity) and the large scale structure tracer are biased tracers of it. 

\begin{equation}
\label{final_cross_correlation}
    \omega_{J,r}(z) = \frac{dJ_{Y}}{dz} b_r b_{im} \omega_{M,M}(z).
\end{equation}

As both the tracer catalog and the EBL photons are biased tracers of the underlying matter field, we include scale-independent linear bias factors $b_{im}$ and $b_r$. This defines the angular cross correlation function, $\omega_{J,r}$ between the tracer catalog and the intensity map, which is the observed quantity. The use of the angular correlation function to extract redshift dependent quantities in this way is due to \citet{Newman_2008}. The explicit result in Equation~\ref{final_cross_correlation}, however, was first arrived at in \citet{mnard2013clusteringbased} who take this as the definition of the angular cross correlation function $w_{J,r}$ between the specific intensity (or its band averaged counterpart) and the large scale structure tracer.

Equation~\ref{final_cross_correlation} is the intrinsic observable obtainable in a broadband measurement, where $J_Y$ is the band averaged specific intensity (Equation~\ref{Big_J}), which depends on the integrated source rest frame emissivities, $\epsilon_\nu$ (see Section~\ref{sec:RT} for our modelling of these quantities).  The matter angular correlation function $\omega_{M,M}$ can be estimated either numerically with tools like CLASS \citep{Blas_2011} or CAMB \citep{Lewis_2000} or with fitting functions \citep{Maller2005}. The assumption of linear or scale independent biasing is valid on the large scales measured here and has been tested in the context of clustering redshift estimation in \citep{Schmidt2013,Rahman2015}. $dJ_Y/dz$ encodes information about the astrophysics of UV photon production, while the remaining terms encode the structure formation history and underlying cosmology dependence.


We will consider the effect of modeling the reference catalog bias $b_r$ in Section \ref{sec:bias_evolution_error} and infer the evolution of the map bias, $b_{im}$, by modeling it as a double power law in redshift and frequency, with evolution parameters $\gamma_{\nu}$ and $\gamma_{z}$: 

\begin{equation}
    \label{photon_bias} 
    b_{im}(\nu, z) = b_{1500}^{z=0} \left(\frac{\nu}{\nu_{1500}}\right)^{\gamma_\nu} (1+z)^{\gamma_z}\,.
\end{equation}
These model the evolution of the frequency and redshift dependent photon clustering bias $b_{im}(\nu, z) \approx b_{im}(\bar{\nu}, z)$ that we evaluate at the effective frequency ($\bar{\nu}$) of the filter in estimating cosmic or sample variance on a per filter basis. The effective frequency we compute averages over the filter response, estimating the effective bias by its average value in the filter. Although choices in how this average is performed may change the estimate of the filter-specific cosmic variance, this is dominated by the bias weighted integral of the correlation function of the underlying matter field, which on the scales we consider is orders of magnitude smaller than the contributions of field to field variance, shot noise in the spectroscopic catalog, and evolution of the bias. As such, the choice to include or exclude filter-specific cosmic variance does not change our forecasted constraints on UVB model parameters.

\section{Emissivity Model} 
\label{sec:RT}

\subsection{Restframe Model}
\label{sec:emissivity}

The comoving UV emissivity-- the frequency-dependent energy emitted per unit time and volume-- is written as the sum of UV photon contributions from all sources. UV photons can be produced by stellar populations in galaxies (written as $\epsilon_{\nu}^{\star}$), by active galactic nuclei ($\epsilon_{\nu}^{\rm AGN}$), and through recombinations ($\epsilon_{\nu}^{\rm rec}$), all of which are considered in the source rest frame. Therefore, the total restframe emissivity $\epsilon_{\nu}$ determined from broadband observations is
\begin{equation}
\epsilon_{\nu} = \epsilon_{\nu}^\mathrm{AGN} + \epsilon_{\nu}^\mathrm{\star} + \epsilon_{\nu}^\mathrm{rec}\,.
\end{equation}
Our model for $\epsilon_{\nu}$ follows \cite{Chiang_2019} and parameterizes the model in \citet{2012ApJ...746..125H}. This model has been compared to broadband tomographic constraints from {\small GALEX} and used to inform improved synthesis modeling in \citet{Faucher_Gigu_re_2020}. We approximate the spectral energy distribution of the EBL over the wavelength range $500$ to $5500$ \AA~as a series of piecewise defined power laws. The filter width of the instrument is much wider than the emission feature, so the Lyman-$\alpha$ emission line can be represented by a delta function at $1216$ \AA. 

For ionizing photons with $\lambda < 912$~\AA, $\nu > 3.29 \times 10^{15} Hz$, we write

\begin{equation}
\label{emiss_less_912}
\epsilon_{\nu} = f_{\rm LyC}(z) \left(\frac{\nu_{1216}}{\nu_{1500}}\right)^{\alpha_{1500}} \left(\frac{\nu_{912}}{\nu_{1216}}\right)^{\alpha_{1216}}\epsilon_{1500} \left( \frac{\nu}{\nu_{912}}\right)^ {\alpha_{912}},  
\end{equation}
where $f_{\rm LyC}$ is a function that parameterizes the evolution of the Lyman continuum escape fraction. We normalize $f_{\rm LyC}$ at redshifts where the Lyman break is directly constrained as it passes through the {\small GALEX} filters, such that
\begin{equation}
\begin{split}
    &\log f_{\rm LyC}(z) = \left( \left( \log{f_{\rm LyC}^{z=2}} - \log {f_{\rm LyC}^{z=1}}\right) / \log{\left(\frac{1+2}{1+1}\right) }\right) \times\\ 
    & \log{ \left(\frac{1+z}{1+1}\right)} + \log{f_{LyC}^{z=1}}.
\end{split}
\end{equation}

$\epsilon_{1500}$ will appear in each expression and normalizes the total emissivity to its value at $1500$ \AA, while $\nu_{\rm X}$ is the frequency corresponding to X \AA. Similarly, $\alpha_{\rm X}$ is the spectral slope at X \AA~for X $= 912$, $1216$ and $1500$ \AA. The spectral slopes at $1216$ and $1500$ \AA~evolve with redshift according to

\begin{equation}
    \alpha_{X} = \alpha_{X}^{z=0} + C_{\alpha X} \log(1+z),
\end{equation}
where $X = 1216, 1500$, and the $\alpha_{X}^{z=0}$ are the values of the power law parameters at $z = 0$ as determined from the {\small GALEX} intensity photometric intensity maps in \cite{Chiang_2019}.

For photons overlapping with Lyman-$\alpha$ emission, that is 912-1216 \AA, or $\nu = 2.47-3.29\times 10^{15}$ Hz, we supplement the power law with a delta function for Lyman-$\alpha$:
\begin{align}
\label{emiss_912_1216}
\epsilon_{\nu} &= \epsilon_{1500} \left(\frac{\nu_{1216}}{\nu_{1500}}\right)^{\alpha_{1500}}  \left[\left(\frac{\nu}{\nu_{1216}}\right)^{\alpha_{1216}}  + D(z, \nu)\right]\nonumber \\
D(z,\nu) &= {\rm EW}_{\rm Ly\alpha}(z) \delta (\nu - \nu_{1216}) \left(\frac{\nu^2}{c} \right). 
\end{align}

Motivated by the midpoints of the {\small GALEX} filter bands, we follow \cite{Chiang_2019} and model the Ly$\alpha$ equivalent width, ${\rm EW}_{\rm Ly\alpha}(z)$, as linear in $\log(1+z)$:


\begin{align} 
{\rm EW}_{\rm Ly\alpha}(z) = C_{Ly\alpha} \log (1+z) + EW_{Ly\alpha}^{z=0.3} \nonumber \\ 
C_{Ly\alpha} = (EW_{Ly\alpha}^{z=1} - EW_{Ly\alpha}^{z=0.3}) / \log \left(\frac{1 + 1}{1+0.3} \right)
\end{align} 

{\small CASTOR} measures redder wavelengths than {\small GALEX}  ($0.15-0.55 \mu$m instead of $0.1-0.15 \mu$m), and so the new data will not measure EW$_{\rm Ly\alpha}^{0.3}$. We therefore fix it to the fiducial value of -6.17 from {\small GALEX}. Similarly, we fix $\alpha_{912} = -1.5$ as this parameter lacks a direct data constraint in {\small CASTOR} or {\small GALEX} photometry. The final piece of the emissivity model is the non-ionizing or long wavelength continuum for photons with wavelengths greater than 1216 \AA~ or $\nu < 2.47 \times 10^{15}$ Hz,
\begin{equation}
\label{emiss_1500}
\epsilon_{\nu}(z) = \epsilon_{1500} \left( \frac{\nu}{\nu_{1500}}\right)^{\alpha_{1500}}\,,
\end{equation}
with redshift evolution
\begin{equation}
    \epsilon_{1500} = \epsilon_{1500}^{z=0} (1+z)^{\gamma_{\epsilon 1500}}.
\end{equation}
Intensity mapping experiments measure a biased tracer of the underlying matter distribution. $\epsilon_{1500}$ is thus constrained only as a product with the $z=0$ bias normalized $b_{1500}^{z=0}$. 

In summary, our emissivity model has 11 free parameters that we evaluate in our inference from $\lambda$ = 700 to 3000 \AA. Four parameters model the power law slope of the emissivity with redshift ($\alpha_{1216}^{z=0}$, $\alpha_{1500}^{z=0}$, $C_{\alpha 1216}$, $C_{\alpha 1500}$). Two model the log of the Lyman continuum escape fractions ($\log{f_{LyC}^{z=2}}, \log{f_{LyC}^{z=1}}$). Additionally, we have the ${\rm Ly\alpha}$ equivalent width at $z=1$ (EW$_{\rm Ly\alpha}^{z=1}$), the product of the emissivity and the bias normalization $\log{(\epsilon_{1500}^{z=0} b_{1500}^{z=0})}$ and the redshift evolution of the emissivity $\gamma_{\epsilon 1500}$. 



\subsection{Projection in Redshift Space} 

The piecewise model of Section~\ref{sec:emissivity} is related to the observed frame intensity $j_\nu$ via Equations~\ref{cosmo_transfer}-\ref{Big_J}.
To derive the observed frame quantity, we also require an instrumental response function which characterizes the transmission fraction or the probability of detecting an incident photon. This yields an instrumental magnitude of a source, which is a combination of the distribution of emitted photons, the distance to the source, and the instrumental response to detected photons. We can thus convert the observed broadband intensity in frequency space into a broadband intensity distribution in redshift. 
Combining Equations~\ref{specific_emiss}-\ref{Big_J} above yields the observed frame quantity desired in Equation~\ref{final_cross_correlation} for a tomographic survey,
\begin{equation}
\label{forward_model}
\begin{split}
\frac{dJ_Y}{dz} b_{im}(z) &= \frac{c}{4\pi H(z) (1+z)}  \\ 
&\int \frac{d\nu_{obs}}{\nu_{obs}} R(\nu_{obs}) b_{im}(\nu, z) \epsilon_\nu(z) e^{-{\tau_{eff}(\nu,z)}},
\end{split}
\end{equation} 
where $c$ is the speed of light, $H(z)$ is the Hubble function, $\tau_{eff}$ is the effective optical depth, $\nu$ is the emission rest frame frequency and $\nu_{obs}$ is the observed frame frequency.  

Equation~\ref{forward_model} is a band averaged intensity distribution and a function only of redshift. In this sense, the instrumental response function has turned the rest-frame emissivity function into a photon distribution in redshift by convolving a frequency and redshift dependent quantity with a function that is of frequency only.  The behavior of this function is most easily seen by considering a single delta emissivity function in frequency that is produced at a range of wavelengths. In this case, one would simply recover exactly the filter curve in redshift space \citep[Figure 1 of ][]{Chiang_2019}. The redshift distribution functions are then filter specific combinations of the evolution of the underlying emissivity distribution and the instrument response.

\section{The CASTOR Filters and Error Budget}
\label{sec:error}

\begin{figure}
\includegraphics[width=0.485\textwidth]{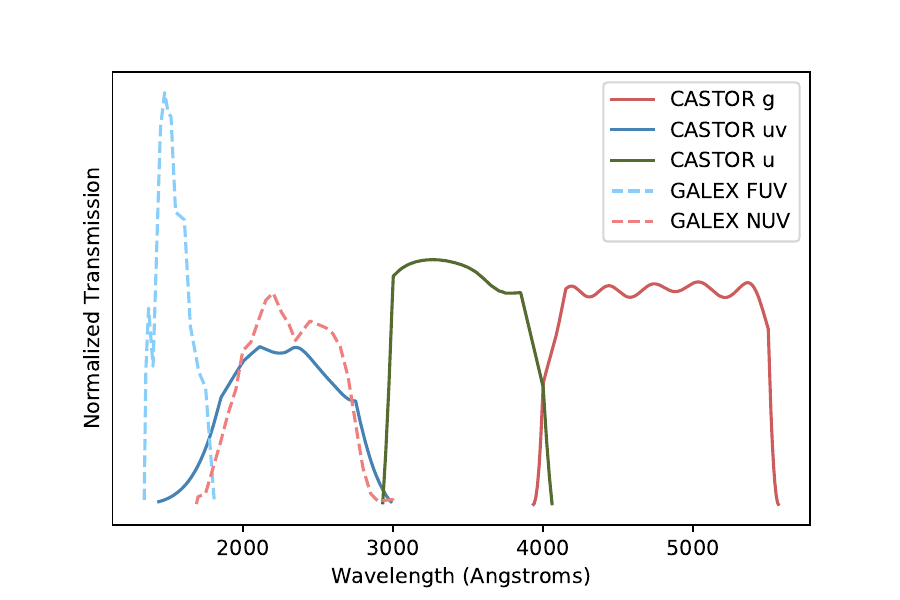}
\caption{\label{fig:filter_curves} The filter transmission curves for {\small CASTOR} UV, u, and g filters (solid) and {\small GALEX} NUV and FUV (dashed). The short wavelength {\small GALEX} FUV filter is not replicated in {\small CASTOR}, which replaces it with two redder filters.}
\end{figure}

{\small CASTOR} is a proposed near UV-optical survey telescope. Wavelength coverage for the UV imaging instrument is $\approx 550-5500$ \AA. The filter response functions  $R(\nu_{obs})$ are shown in Figure~\ref{fig:filter_curves}, where we have also included the {\small GALEX} $\{NUV, FUV\}$ filter sets for comparison. {\small GALEX} covers effective wavelengths $0.1-0.15 \mu$m in the FUV and NUV, compared to effective wavelengths of $0.23-0.5 \mu$m for the $\{uv, u, g\}$ filters on {\small CASTOR}. The NUV and uv filters on {\small GALEX} and {\small CASTOR} provide similar constraining power, while the $u$ and $g$ filters extend the data constraints in observed frame frequency and thus higher redshift. {\small CASTOR} thus has weaker constraints on short wavelength emission at low redshift, but greater potential to constrain shorter wavelengths at higher redshifts in the red filters. While GALEX samples the continuum to $z = 1$ in the FUV and $z=2$ in the NUV, {\small CASTOR}'s redder filters extend these constraints to $z=2.5-3$ in the $g$ band. 
 
In this section we will describe our error models for a {\small CASTOR} like survey. We will consider the contributions from shot noise (Section~\ref{sec:shotnoise}),  photometric zero point (Section~\ref{sec:zeropoint}), and evolution of the reference catalog bias (Section~\ref{sec:bias_evolution_error}). We place upper and lower bounds on the total error budget from these contributions.

\subsection{Shot Noise} 
\label{sec:shotnoise}

For a galaxy tracer-intensity map cross correlation, shot noise is introduced due to both the discrete nature of the galaxy tracers and the contribution from the tracers to the observed intensity in the map. In other words, for a flux weighted cross correlation, the amplitude of the shot noise becomes flux weighted. To estimate the size of the shot noise on estimates of the correlation function, one can use a counts in cells approach, as in \citet{Wolz_2017} developed for HI intensity mapping or work directly from the definition of the cross correlation estimator. In either case, for a tracer with angular number density $n_g$, the variance in flux weighted counts is then (cf. \citealt{gabrielli2006statistical}),

\begin{equation}
    \sigma_{SN}^2 = \frac{\left<J_Y\right>}{n_g}.
    \label{eqn: flux_weighted_counts}
\end{equation}

This corresponds to a scale independent shot noise with amplitude set by the angular number density of tracer objects per redshift bin, $n_g$. In general, the tracer catalog density will be a function of both the galaxy density function in units of number per steradian, which can be computed from the halo mass function dN/dM, where dN is the number per mass bin M+dM, by integrating in mass and over the cosmological volume and a survey selection function.


To model the tracer distribution as a function of redshift, we take the SDSS CMASS and LOWZ \citep{Reid2016}, eBOSS LRG and QSO \citep{Ross2020}, and SDSS QSO DR12 \citep{Paris2017} and DR14 \citep{Paris2018} catalogs and divide the redshift distribution into $80$ bins from redshift $0-4$. In total, this corresponds to 2,727,612 tracer objects distributed over $\approx 7000$ square degrees in the northern hemisphere. 

\begin{figure}
\resizebox{8.5cm}{!}{\includegraphics{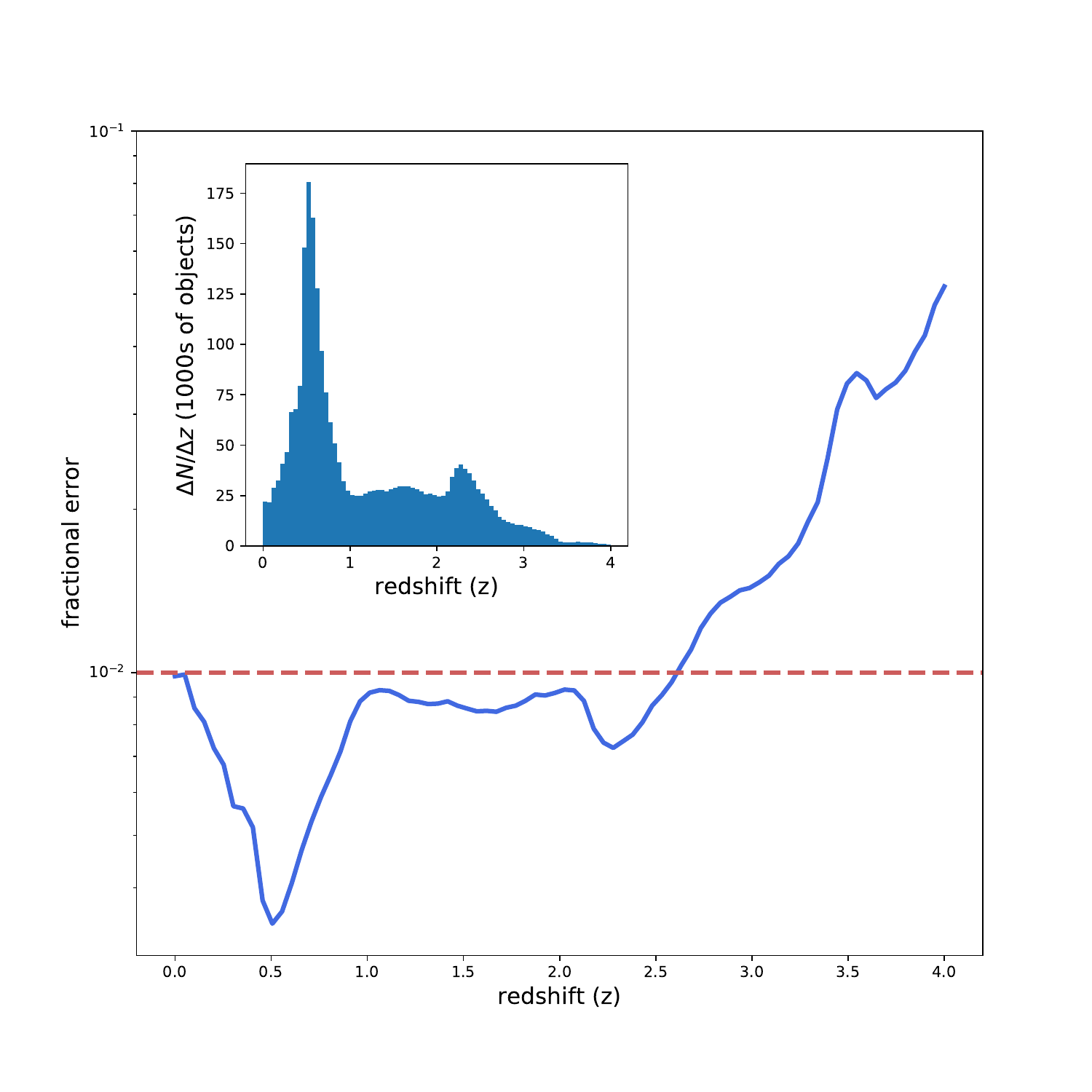}}
\caption{\label{fig:Model1} The fractional error budget as a function of redshift $\sim$ $1/\sqrt{n_g}$. Also plotted is a photometric zero point error for which we assume a fixed $1\%$ value in our optimal error model. Inset is the redshift distribution from the SDSS tracer surveys we consider in this work. The corresponding shot noise curve is multiplied by a factor of 5 to account for the improved depth of future spectroscopic catalogs, and calculated using the relationship derived in \S~\ref{sec:shotnoise}}
\end{figure}

We plot the total error budget in Figure~\ref{fig:Model1}. At most redshifts, the photometric zero point error is larger than the shot noise component. Future spectroscopic tracer catalogs are expected to improve on the completeness, that is, detect a larger fraction of galaxies contributing to the diffuse component of the UVB, of existing surveys, especially at $z>1$. In Section~\ref{sec:zeropoint} we discuss how uncertainty in the field to field photometric zero point propagates to a $1-3 \%$ uncertainty in the angular correlation function. Shot noise from existing tracer catalogs is subdominant to the photometric error for $z \gtrsim 0.2$ and for $z \lesssim 2.6$.  In Section~\ref{sec:error_summary} we will consider two noise models that bound the uncertainty in the recovery of the intensity distribution in redshift. In order to set the lower bound, we consider a spectroscopic tracer catalog improved in depth, as measured by the number density of sources in each redshift bin, by a factor of 5, which delivers a shot noise term that is subdominant to the photometric error for $z<4$. Concurrent with the nominal lifetime of the observatories we consider, in the mid to late 2020s, several large scale structure experiments will deliver deeper spectroscopic catalogs. In particular, the Dark Energy Spectroscopic Instrument is expected to deliver a factor of 10 improvement in depth over the catalogs we consider, $\approx 30$ million galaxy and quasar redshifts, between $z=0$ and $z \gtrsim 3.5$ \citep{collaboration2016desi}.

The shot noise estimate given in Equation~\ref{eqn: flux_weighted_counts} assumes that noise is due entirely to emission from discrete sources. That is, we neglect the variance in estimating the flux from an intensity map discretized due to the finite resolution of an observed intensity map. This underestimates the error since the intensity map will also have contributions from diffuse emission due to extra-galactic sources below the detection limit of spectroscopic tracer catalogs. Given the deeper spectroscopic catalogs we consider, we expect the overall effect of neglecting the contributions of diffuse sources to be a larger effect for current than for future surveys. This leads us to expect that, with the improved depth of future spectroscopic catalogs, neglecting the diffuse extra-galactic component in our analytic approach will lead to only a small underestimate of the total noise amplitude. 

It is important to note that our definition of the map bias, $b_{im}$, encodes that the observed map intensity $J_{Y}$ is a biased tracer of the underlying matter density field. As a result, the estimate of the shot noise in Equation~\ref{eqn: flux_weighted_counts} is a bias weighted estimate. Since dim and diffuse sources will be less clustered than more massive sources that produce a greater observed intensity, we expect that the contribution of diffuse sources to the shot noise in the map will be less than the contribution from the discrete sources considered in the cross-correlation. 


The frequency resolution for discrete features of the spectral energy distribution is set by the ratio between the reference catalog binning $\delta_r$, in redshift space given by $\Delta z = 0.05$, and the intrinsic clustering scale $\delta_c$ \citep{mnard2013clusteringbased}. This is because the clustering scale is a small number when evaluated in redshift space. A $5$ Mpc angular separation corresponds to $\Delta z \approx 0.001$ at $z = 0$ and $\Delta z \approx 0.01$ at $z \approx$ 1. The signal to noise ratio also increases with the effective survey volume, owing to more tracer objects in the reference catalog. Broadband estimates of the SED therefore favor larger survey volumes and hence larger redshift binning.

For existing spectroscopic catalogs, shot noise is comparable to photometric errors. However, improved tracer catalogs reduce uncertainties by a factor of the survey depth, thus requiring improved photometric error control. Still broadband tomography may provide advantages over other measurement techniques as these are often limited by the ability to perform a foreground decomposition into galactic and extragalactic components. 

\subsection{Error due to Bias Evolution in the Tracer Catalog}
\label{sec:bias_evolution_error}

From Equation \ref{final_cross_correlation}, the ability to extract the bias weighted redshift distribution $\frac{dJ_{\nu}}{dz} b_{im}(\nu, z)$ is limited by knowledge of reference catalog bias $b_r$ evolution with redshift.
Uncertainty in the angular correlation function and the tracer catalog bias propagates to the cross-correlation, $\frac{dJ}{dz} b_{im}$ (eq.~\ref{forward_model}). We can estimate the contribution of the bias evolution to the inferred distribution by considering the mean offset in the inferred redshift,
\begin{equation}
    E[\hat{z}] - z = \int z^{\alpha+1} \mathcal{N}(z_0, \sigma) dz - \int z \mathcal{N}(z_0, \sigma) dz,
    \label{eqn:biasevol}
\end{equation}
%
We assume a Gaussian probability distribution, $\mathcal{N}(z_0, \sigma)$ for the emission redshift, $z$. $z_0$ is the true mean redshift of the emission, $\sigma$ is the associated standard deviation of the distribution about $z_0$ and $\alpha$ is a parameter describing the bias evolution such that $b_r \propto z^\alpha$. Following \citet{mnard2013clusteringbased}, we assume that $\alpha = 0.1$, that is, estimates of the bias neglect up to $10\%$ of its evolution with redshift and that $\sigma \approx 0.5$ based on the SDSS redshift distribution. We show the evolution of Equation~\ref{eqn:biasevol} with redshift as the dotted curve in Figure \ref{fig:Error Models}. This effect is largest at low redshift and decreases rapidly at higher redshift. 


\subsection{Photometric Zero Point and Cosmic Variance} 
\label{sec:zeropoint}

Errors in the photometric zero point of the intensity map and tracer catalog contribute to the determination of the angular correlation function \citep{Coil2004} due to the finite field of view measured by each exposure. Varying zero points between fields produce an effectively varying magnitude limit, which in turn produces a difference in map depth and a change in the estimated surface brightness. That is, since we take correlations between the tracer catalog and the absolute overintensity, variations in the effective mean intensity $\left<J_{Y}\right>$ propagate to cause spurious changes in $\Delta J_{Y}$.

A second and smaller effect arises from differences in the photometric zero point, or catalog depth, of the spectroscopic tracer catalog. Here, the effect is to increase the variance beyond the typical Poisson 1/N scaling as each field varies due to the fluctuating magnitude limit. The photometric effect scales with the photometric zero point fluctuation amplitude and, since it is a consequence of field to field variations, the number of fields over which the cross correlation is measured. The spectroscopic effect is an order of magnitude smaller, \citet{Newman_2008} estimated this latter effect at the roughly 0.1$\%$ level in Monte Carlo tests under conservative assumptions.

For \small{CASTOR}, the mirror diameter increases to 1 m from 0.5 m for $\small{GALEX}$ leading to increases in the point-source derived limiting magnitudes from 19.9 fuv, 20.8 nuv, in GALEX All-Sky Imaging Survey (AIS) to $\approx$ 27, in the uv, u, and g for CASTOR respectively \citep{2005ApJ...619L...7M}. Although not derived for a low surface brightness background measurement in a photometric intensity mapping survey, we expect similar corresponding improvements in the photometric error properties for $\small{CASTOR}$. A rough estimate for the photometric error, which scales with mirror area, is that it should improve by a corresponding multiple of the improvement in the mirror size. This is because the field to field variance is due to drop out of sources, and the probability of detecting a source in a given field is a function of the limiting magnitude, set by the integration time and mirror size. The important quantity, here, is not the absolute photometric zero point, but the variation in zero point between fields. Together, the zero point effects were estimated to contribute on the order of $3\%$ to the fractional error of the GALEX intensity maps in \citet{Chiang_2019}. Given both the several magnitude improvement in the point source limiting magnitude and the $4\times$ increase in mirror area with CASTOR, we thus naively expect photometric zero point error to be $\le 0.75$\%.
However, achieving better than $1\%$ error control is always challenging, and so we will conservatively assume a $1\%$ photometric zero point error. Note that the photometric zero point error dominates the error budget. The most worst case would be the $3\%$ photometric zero point error from GALEX, which would achieve similar constraints to \cite{Chiang_2019}. 

We also considered a more general model which allows for the growth of photometric errors with redshift. The first term of Equation~$\ref{forward_model}$ models the dimming of the source at higher redshift. Motivated by the form of this expression, while noting that $J_Y$ and the bolometric flux differ by an additional factor of 1/(1+z), we assume photometric error scales with the flux, which scales with $1/(1+z)^2$. We assume a functional form
\begin{equation}
    \label{eqn: photo_z_growth} 
    \sigma_P = A_n(1+z)^2
\end{equation}
where we take $A_n$, the noise amplitude, to be approximately the same level as the fixed component of the photometric noise, i.e. $1\%$. This has the effect of rescaling the errorbars with redshift such that the distribution becomes photometric noise dominated at all redshifts for existing spectroscopic tracer catalogs. 

On large scales, the correlation function is limited by the finite number of modes available. However, we use the correlation function only on degree scales, while the tracer survey covers $7000$ degrees. We thus expect cosmic variance to contribute only at the $10^{-6}$  level, negligible in comparison to shot noise and photometric uncertainties \citep{Moster_2011}. 

\subsection{Summary of Error Models and Optimal Spectroscopic Tracer Catalogs}
\label{sec:error_summary}
We combine the error sources above into two error models, which bound the upper and lower limits of uncertainty in the precision of a {\small CASTOR} measurement.
\begin{figure}
\includegraphics[width=0.485\textwidth]{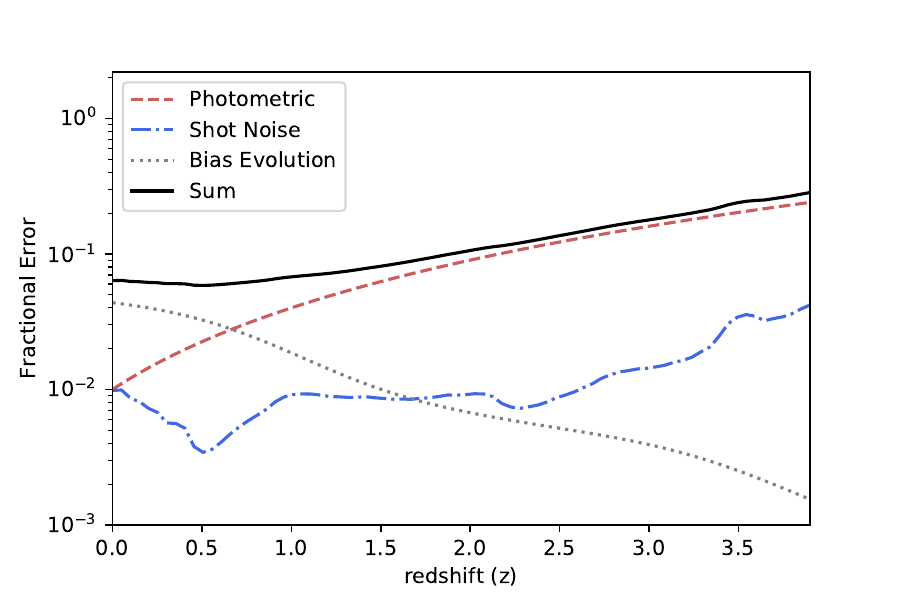}
\caption{\label{fig:Error Models} Fractional Error as a function of redshift for Model C. Our error model consists of three components, a photometric error, shot noise in the spectroscopic tracer catalog and noise related to systematic error in parameterization of the bias evolution with redshift. Model C incorporates each source of error, while Model O sets a lower limit on the errorbars due to shot noise and a fixed photometric zero point error.}
\end{figure}

\begin{enumerate}
    \item Model O: In this model, we consider only shot noise and fixed photometric zero point. This is consistent with quoted errorbars on clustering redshift estimation in simulations and plotted in Figure~\ref{fig:Model1} \citep{Rahman2015}. Since we seek a lower bound on the total uncertainty, we consider an optimal spectroscopic tracer catalog such that shot noise is always less than a fixed photometric error. This optimal catalog is assumed to have five times the depth but the same distribution as the SDSS catalogs discussed in Section~\ref{sec:shotnoise}. For comparison, DESI achieves an $\approx 10$ fold increase in tracer catalog depth and so will achieve optimal tracer density. 
    \item Model C: To the errorbars in Model O, we add the effects discussed in Sections~\ref{sec:bias_evolution_error} and~\ref{sec:zeropoint}, that better reflect the errorbars in broadband tomographic measurements found in \citep{Chiang_2019} and \citep{Chiang20}. We assume only the existing spectroscopic catalogs in modeling the shot noise component. 
\end{enumerate}

We show Model C in Figure~\ref{fig:Error Models}. Model O (whose evolution with redshift is shown in Figure~\ref{fig:Model1}) is a lower bound on the error due to fixed additive noise from varying photometry and tracer catalog completeness, while Model C is likely to overestimate the real low redshift error bars owing to its simple parameterization of weakly constrained bias evolution and assumption that photometric errors increase with redshift rather than remaining constant. The fractional error grows with redshift, but not rapidly enough to produce large absolute uncertainties given the decline in the intensity with redshift. Therefore, the signal and the error together go smoothly to zero at high redshift because of the pre-factor in Equation~\ref{forward_model} and the shape of the filter, regardless of the underlying SED shape that we constrain. 

The conservative error estimates we obtain show that the common assumption of 1/N Poisson noise \citep{Scottez18} tends to underestimate the true uncertainties. Further, comparison of our analytic approach with bootstrapped errorbars from GALEX in \cite{Chiang_2019} reveals an unaccounted for and redshift dependent term that inflates the error by a factor of 2-3 at $z > 0.7$. Another possibility is unmodeled systematics in the reference catalog, owing to differences in the underlying tracer populations. Although we do not attempt to explicitly model this effect, we expect that it is a consequence of the filter shape, where fewer photons are detected as the emission redshifts out of the filter coverage. A stronger evolution of the photometric error could capture this effect at high redshift at the cost of less agreement at lower redshift where most of our constraining power is expected to be.

To summarize the redshift dependencies of each model. In Model O, shot noise is subdominant to photometric uncertainties which become comparable as the completeness of the spectroscopic tracer catalog falls off at $z > 2.5$. In Model C, at low redshift, error due to bias evolution dominates, while for  $z \gtrsim 0.5$, photometric errors grow and dominate over both shot noise and bias evolution.

In the remainder of this work, we will show results from both models. We expect the real performance to lie somewhere in between.

\section{CASTOR Results} 
\label{sec:results}

The bias weighted intensity distribution contains information about both the spectral energy distribution (SED) of the extragalactic background and its overall intensity or monopole term. In Section~$\ref{SED}$ we estimate uncertainties on the parameters governing the shape of the SED model and its evolution. In Section~$\ref{EBL}$ we infer the EBL monopole conditioned on the SED model parameter distribution as a convenient statistic for summarizing measurements of the EBL in a technique independent fashion.

\subsection{UV-Optical Background Spectral Energy Distribution} 
\label{SED}

To estimate {\small CASTOR}'s sensitivity to the parameters of the underlying SED model, we write the likelihood of the fiducial model given a vector of model parameters $\Psi$ as 

\begin{equation}
\label{likelihood}
p(D|\Psi) = \mathcal{N}\bigg(\frac{dJ}{dz}b_{im}(\Psi, z), \sigma^2\bigg)
\end{equation}
where $\sigma^2$ is computed by adding the terms in each error model in quadrature. Here, $\mathcal{N}$ indicates the Gaussian normal distribution. The probability that a vector $\Psi$ defines the true underlying model is given by Bayes' theorem.

\begin{figure*}
    \begin{center}
    \resizebox{15.0cm}{!}{\includegraphics{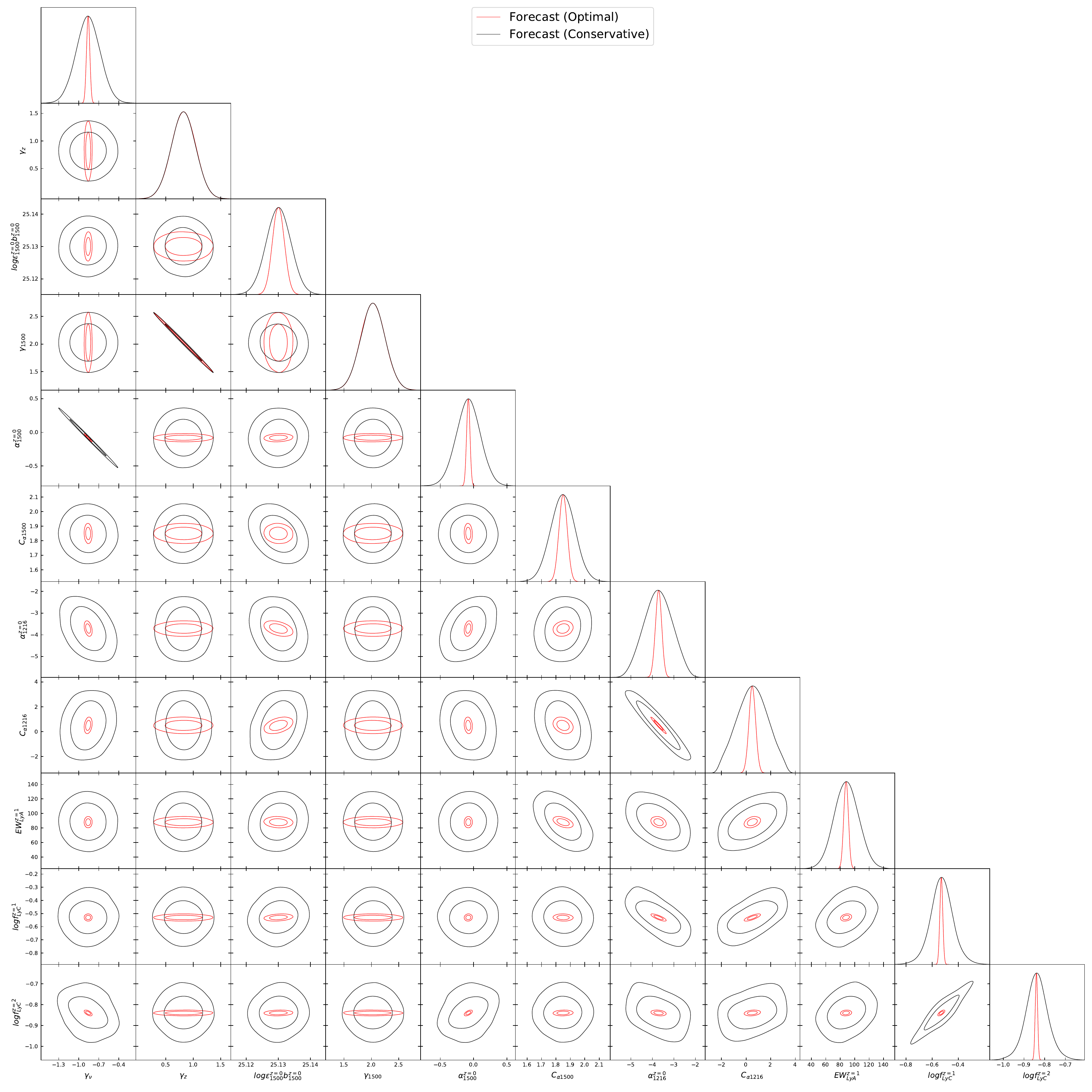}}
\caption{\label{fig:corner} From left to right, posterior distributions for the parameters of the SED model $\gamma_{\nu}$, $\gamma_z$, $\log(\epsilon_{1500}^{z=0} b_{1500}^{z=0})$, $\gamma_{1500}$, $\alpha_{1500}^{z=0}$, $C_{\alpha 1500}$, $\alpha_{1216}^{z=0}$, $C_{\alpha 1216}$, EW$_{Ly\alpha}^{z=1}$, $\log f^{z=1}_{LyC}$, $\log f^{z=2}_{LyC}$. Red contours indicate uncertainties for the optimal error model using a spectroscopic tracer catalog with five times the depth of the SDSS and a fixed photometric uncertainty. Black curves indicate corresponding uncertainties in a conservative error model which adds a redshift dependent photometric uncertainty, bias evolution, and shallower tracer catalog. Diagonal panels show marginalized posteriors for each parameter, while off-diagonal panels show the relationships between model parameters. The geometric mean improvement of the optimal forecast over the conservative model is a factor of $\approx$ 5 and a factor of $\approx 10$ better than GALEX. As discussed in the text, $\log{\epsilon_{1500}b_{1500}}$, $\gamma_z$ and $\gamma_{1500}$ are prior dominated or see minimal improvements with the additional and redder filter coverage.}
    \end{center}
\end{figure*}

We assign uniform priors on each parameter with width specified by the marginalized posterior uncertainties in \cite{Chiang_2019}. We sample our likelihood function using the affine-invariant Markov Chain Monte Carlo code, {\small EMCEE} \citep{Foreman_Mackey_2013}. Table~\ref{tab:table1} summarizes our fiducial model parameters and their corresponding priors. As mentioned in Section~\ref{sec:emissivity}, we do not constrain ${\rm EW}_{\rm Ly\alpha}^{z=0.3}$, since we lack the FUV filter present on GALEX and thus cannot directly observe Ly$\alpha$ emission at low redshift. As the additional and overall redder filter set on {\small CASTOR} is not expected to improve constraints on the overall amplitude of the bias weighted intensity normalization at 1500 $\mathrm{\AA}$ compared to the constraints from {\small GALEX}, we place Gaussian priors on $\log(\epsilon_{1500} b_{1500})$, the photon clustering bias evolution with redshift $\gamma_z$ and the evolution of the 1500 $\mathrm{\AA}$~normalization $\gamma_{\epsilon 1500}$. 

\begin{table}
  \begin{center}
  \caption{Priors on and fiducial values for parameters of the SED model.\label{tab:table1} \textbf{as derived from the best fit parameters and uncertainties in GALEX constraints on the UVB.}}
    \begin{tabular}{c c c c}
      \textbf{Parameter} & \textbf{Range} & \textbf{Type} & \textbf{Fiducial}\\
      \hline
      $\gamma_{\nu}$ & [-3.44, +0.8]& Flat & -0.86\\ 
      $\gamma_{b}$ & $\sigma$ = 0.33 &  Gaussian & 0.79\\ 
      $\log(\epsilon_{1500} b_{1500}^{z=0})$ & $\sigma$ = 0.01 & Gaussian & 25.13 \\
      $\gamma_{1500}$ & $\sigma$ = 0.30 & Gaussian & 2.06\\
      $\alpha_{1500}^{z=0}$ & [-1.76, 2.48] & Flat &  -0.08\\
      $C_{\alpha 1500}$ & [-0.71, +4.29] & Flat & 1.85\\
      $\alpha_{1216}^{z=0}$ & [-5.67, -1.03] & Flat & -3.71 \\
      $C_{\alpha 1216}$ & [-2.38, 3.42] & Flat & 0.5\\
      EW$_{Ly\alpha}^{z=1}$ & [-9.72, 190.9] & Flat & 88.02\\
      $\log f^{z=1}_{LyC}$ & $<$0 & Flat & -0.53\\
      $\log f^{z=2}_{LyC}$ & $<$0 & Flat & -0.84\\
    \end{tabular}
  \end{center}
\end{table}

\begin{figure}
\resizebox{8.5cm}{!}{\includegraphics{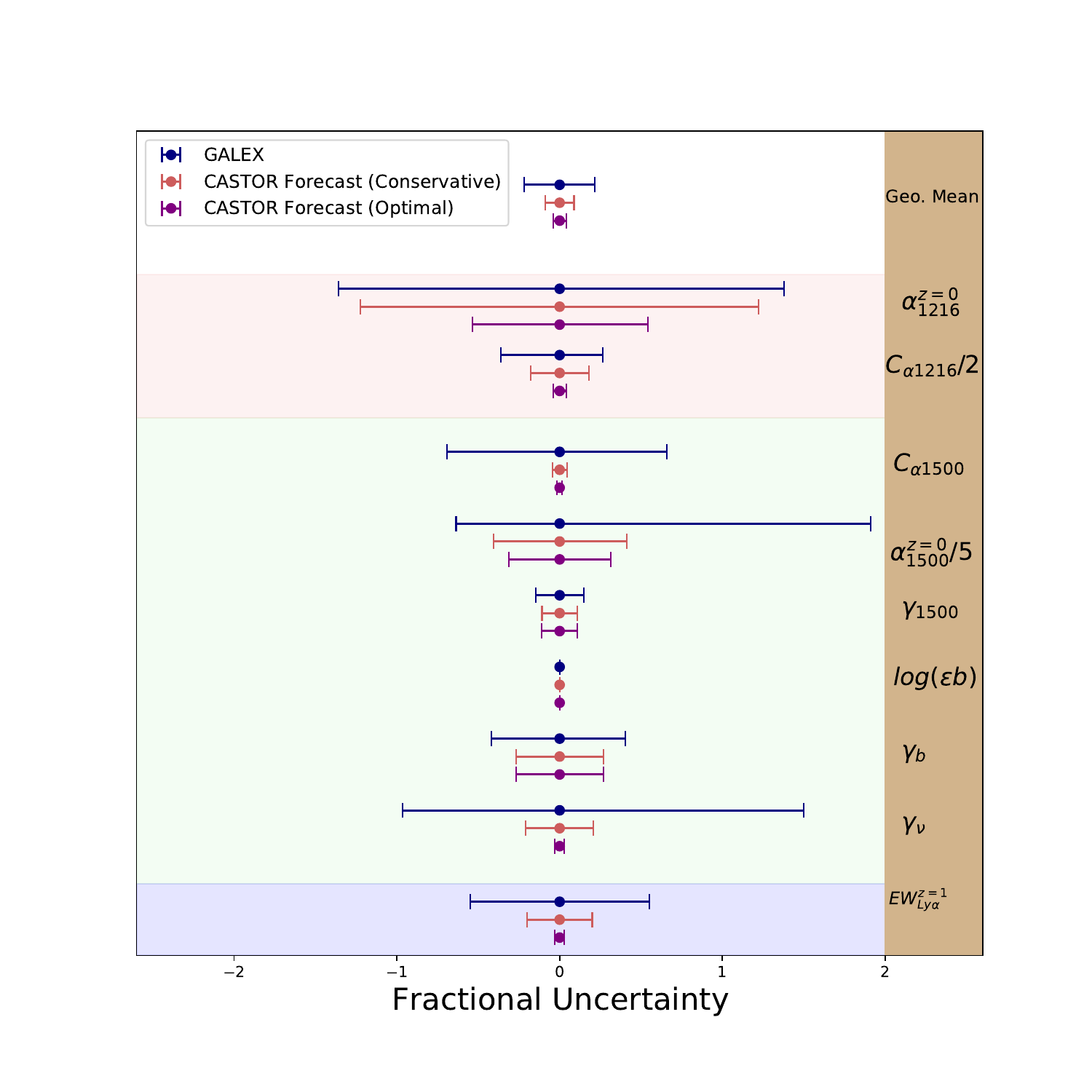}}
\caption{ \label{SED_comparison} Comparison of the posterior fractional uncertainties on the SED model parameters forecasted for {\small CASTOR} conservative (red) and optimal (purple) error model to the constraints from GALEX data (blue) in \citep{Chiang_2019}. Parameters in the red region are constrained by the data to $z \approx 4$, in the green region to $z \approx 2.7$, and in the blue region at $z = 1$. Also shown is the geometric mean of the fractional uncertainties for each.} 
\end{figure}

\begin{figure*}
    \begin{center}
    \includegraphics[width=1.25\textwidth, center]{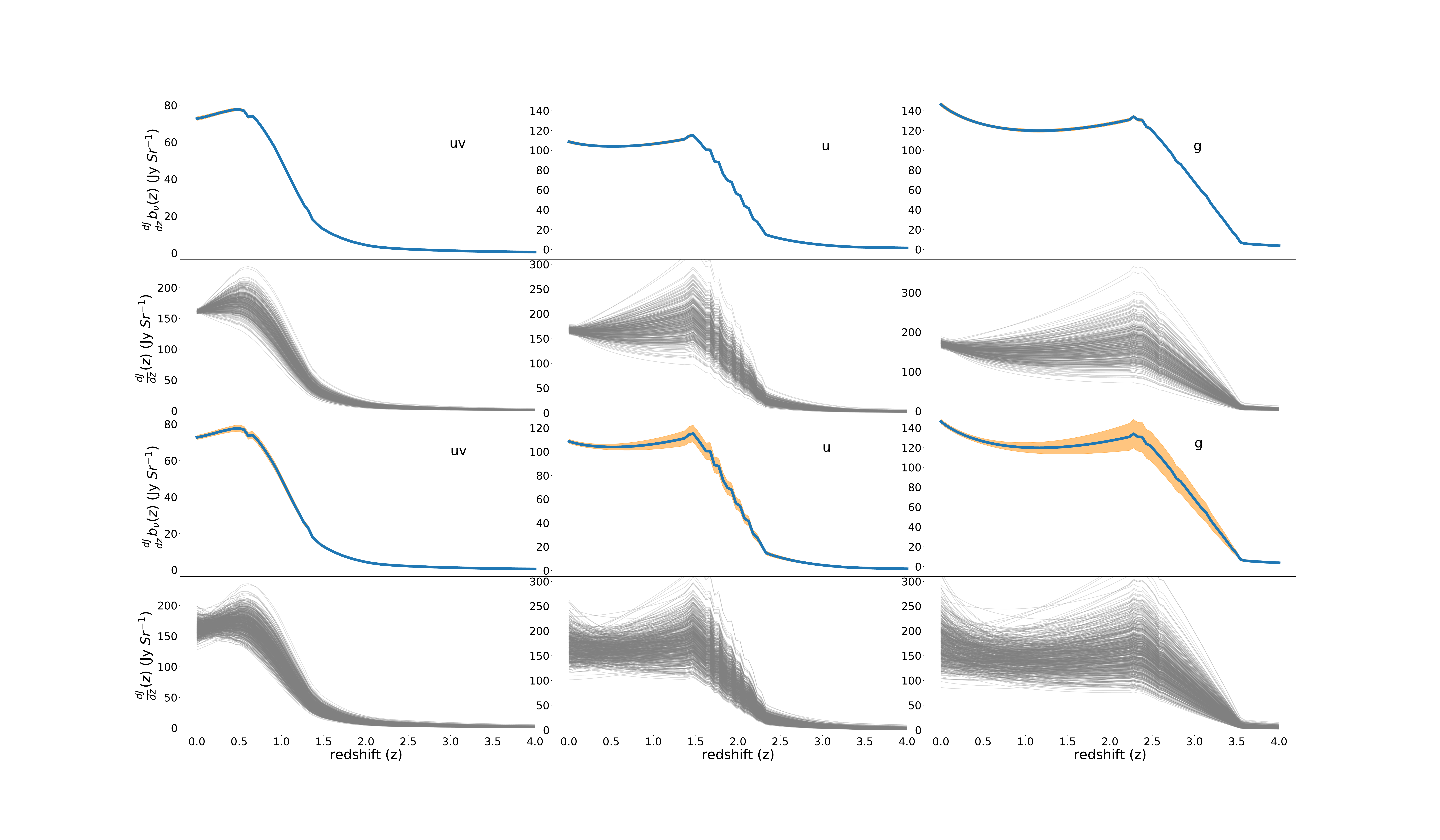}
    \end{center}
    \caption{\label{fig:dJdzbjuv} First row: The bias weighted specific intensity distribution, $\frac{dJ}{dz} b_{im}(z)$, as a function of redshift for the {\small CASTOR} uv imager filters. The shaded region represents the error budget as a function of redshift, determined from our optimal tracer catalog and fixed photometric error. Second row: Intensity distribution in redshift with bias removed and sampling of corresponding fits to the distribution from the SED posteriors. Also indicated is the magnitude of the filter specific EBL monopole. Third row: Same as first row for our second error model incorporating redshift dependent photometric errors, bias evolution, and a combination of existing spectroscopic tracer catalogs. Fourth row: Same as the second row but for the second error model.}
\end{figure*}


Figure~\ref{fig:corner} shows the 2D contours and 1D marginalized uncertainties for our fiducial model conditioned on the forecasted bias weighted intensities under each error model. These values are summarized in Table~\ref{tab:table2}. Our uncertainties are given as the $67\%$ interquartile range relative to the fiducial model parameter. 

The parameters governing the evolution with redshift of the frequency evolution and the $1500 \AA$\ evolution, $\gamma_b$ and $\gamma_{\epsilon 1500}$, are coupled because they are only measured via the evolution of the SED amplitude, $\log(\epsilon_{1500} b_{1500})$. Similarly, in the frequency evolution of the normalization, there is a degeneracy between $\alpha_{1500}$, the evolution of the emissivity with frequency and the frequency dependent clustering bias evolution $\gamma_{\nu}$. Since {\small CASTOR} NUV is only sensitive to wavelengths greater than rest $1500$ \AA, a similar degeneracy between the $C_{\alpha 1216}$ and $\alpha_{1216}^{z=0}$ parameters arises. A change in the slope can offset a change in the intercept of the evolution of the emissivity in frequency. Finally, the LyC escape fraction is parameterized by differences in the data constraints from the {\small GALEX} filters at $z=1$ and $z=2$, leading to degeneracies between the inferred values of $\log f_{\rm LyC}^{z=2}$ and $\log f_{\rm LyC}^{z=1}$. 




\begin{table}
  \begin{center}
        \caption{\label{tab:table2} Posteriors interquartile ranges for parameters of the SED model under a conservative and optimal error model. The upper and lower limits are the 67$\%$ interquartile range. }
    \begin{tabular}{c c c}
      \textbf{Parameter} & \textbf{Optimal} & \textbf{Conservative} \\
      \hline
      $\gamma_{\nu}$ & [-0.88, -0.83]& [-1.04, -0.68] \\ 
      $\gamma_{z}$ & [0.60, 1.04] &  [0.60, 1.04] \\ 
      $\log(\epsilon_{1500}^{z=0} b_{1500}^{z=0})$ & [25.128, +25.131] & [25.126, +25.133]  \\
      $\gamma_{\epsilon 1500}$ & [1.81, 2.25] & [1.81, 2.24] \\
      $\alpha_{1500}^{z=0}$ & [-0.1, -0.06] & [-0.26, 0.1] \\
      $C_{\alpha 1500}$ & [1.82, 1.85] & [1.77, 1.93] \\
      $\alpha_{1216}^{z=0}$ & [-3.85, -3.7] & [-4.34, -3.06] \\
      $C_{\alpha 1216}$ & [0.23, 0.77] &  [-0.77, 1.8] \\
      EW$_{Ly\alpha}^{z=1}$ & [84.90, 91.20] & [71.9, 105.6] \\
      $\log f^{z=1}_{LyC}$ & [-0.54, -0.52] & [-0.61, -0.44]\\
      $\log f^{z=2}_{LyC}$ & [-0.85, -0.83] & [-0.89, -0.79]\\
      Monopole (uv) Jy/Sr & [300, 403] & [300, 405] \\
      Monopole (u) Jy/Sr & [180, 220] & [180, 221]\\
      Monopole (g) Jy/Sr & [384, 547] & [382, 562]\\
    \end{tabular}
  \end{center}
\end{table}

We compare the relative uncertainties from {\small GALEX} and both {\small CASTOR} error models, along with the range under which that parameter is constrained by the data, in Figure~\ref{SED_comparison}. These represent competitive or improved  uncertainties over those quoted for {\small GALEX} with a geometric mean improvement of a factor of 2-3 for the conservative error model and a factor of $\approx 10$ for the optimal error model. 

In particular, the improved constraints on the $1216$ \AA\ continuum slope, which are unconstrained by {\small GALEX}, are driven by the deeper redshift coverage of a {\small CASTOR}-like survey. Constraints on the parameters of the spectral energy distribution provide additional windows into the ionization and thermal history of the IGM out to $z \approx 3$ that can be compared with constraints at higher redshift.

$\log f_{\rm LyC}$ at $z=1$ and $z=2$ has a $10-20\%$ uncertainty in the conservative error model, and $\approx 1\%$ uncertainty in the optimal model. These constraints offer a window into the UV photon production history at intermediate redshifts. Previous measurements of the total photon budget in the UV favour models for reionization with significant contributions from galaxies, but require large escape fractions \citep{Finkelstein2012}. Both observational and theoretical measurements of the escape fraction of ionizing photons from galaxies are poorly constrained. Observational measurements of the escape fraction from $z\approx 3$ galaxies find low escape fractions ($< 10$ percent) that cannot account for galaxy-driven reionization \citep{siana10,siana15} and escape fractions derived from simulations find wildly varying results \citep{gnedin08,wise09,anderson17}. However, if a larger proportion of ionizing photons escape from low luminosity galaxies at high redshift, then the low measured escape fraction in low redshift galaxies can be reconciled with galaxy driven reionization \citep{Escude,McQuinn_2013}. Although our approach in constraining the escape fraction in this work does not distinguish between diffuse and discrete components, instead modeling the total combined spectral energy distribution, measurements with real data can be decomposed and considered separately by masking sources. This then yields a direct integrated constraints on the sources of diffuse photons to $z \approx 3$.

SED information is captured by differences in the relative filter responses as a function of redshift. In Section~\ref{sec:shotnoise} we considered the contribution from deeper tracer catalogs with fixed source distributions as well as improvements in the photometric uncertainties. Since the tracer catalog, and hence shot noise, is redshift dependent, we expect changes in the inferred parameter uncertainties to be driven more by improvements in the spectroscopic tracer catalog than by uniform multiplicative improvements in the photometric noise. We tested this by running additional test chains with a flat error model independent of redshift and one that was a linear function of redshift, finding modest improvements for the latter over the former at fixed mean uncertainty.

In summary, improved uncertainties in model parameters are driven both by the information present in the additional filters and reduction in the cross correlation error bars. The addition of a third filter both appends a third column to our data vector and extends the constraints to higher redshift. Parameters normalized to their evolution at 1216 \AA\ are constrained to $z\approx 3.5-4$, while parameters normalized at 1500 \AA\ are constrained to $z\approx 2.7$. This compares to limiting redshifts on the data constraints of $z<1$ for {\small GALEX} FUV/NUV or the {\small CASTOR} uv filter alone.

\subsection{Total EBL Monopole} 
\label{EBL}

The EBL monopole is the leading order contribution to the spherical harmonic decomposition of the EBL. This makes it a convenient summary statistic at a given effective frequency for comparing EBL constraints across a variety of measurement techniques and frequency ranges.  Further, beyond being a summary statistic, the EBL monopole intensity at a given effective frequency includes information about a combination of astrophysical emission and cosmic structure \citep{Hill2018}. 

The EBL monopole is determined from the bias weighted intensity distribution functions, $\frac{dJ}{dz}b_{im} (z)$, shown in Figure~\ref{fig:dJdzbjuv} for both error models by integrating Equation \ref{forward_model} over redshift from $z = 0$ to $z = z_{max}$ and fixing the redshift dependent photon clustering bias to its fiducial value as determined from the best fit model. Although the photon clustering bias is only measured as a product with the emissivity normalization, it can be obtained in regions where the frequency evolution of the bias is known to be flat and where a discrete source catalog exists \citep{Chiang_2019}. Estimated monopole values and uncertainties are given in Table~\ref{tab:table2}. Error bars are determined from $67\%$ inter-quartile ranges on the monopole values determined from the posteriors to the parameter fits, a sampling of which are shown in the lower panels of Figure~\ref{fig:dJdzbjuv}. 

\begin{figure}
\resizebox{8.5cm}{!}{\includegraphics{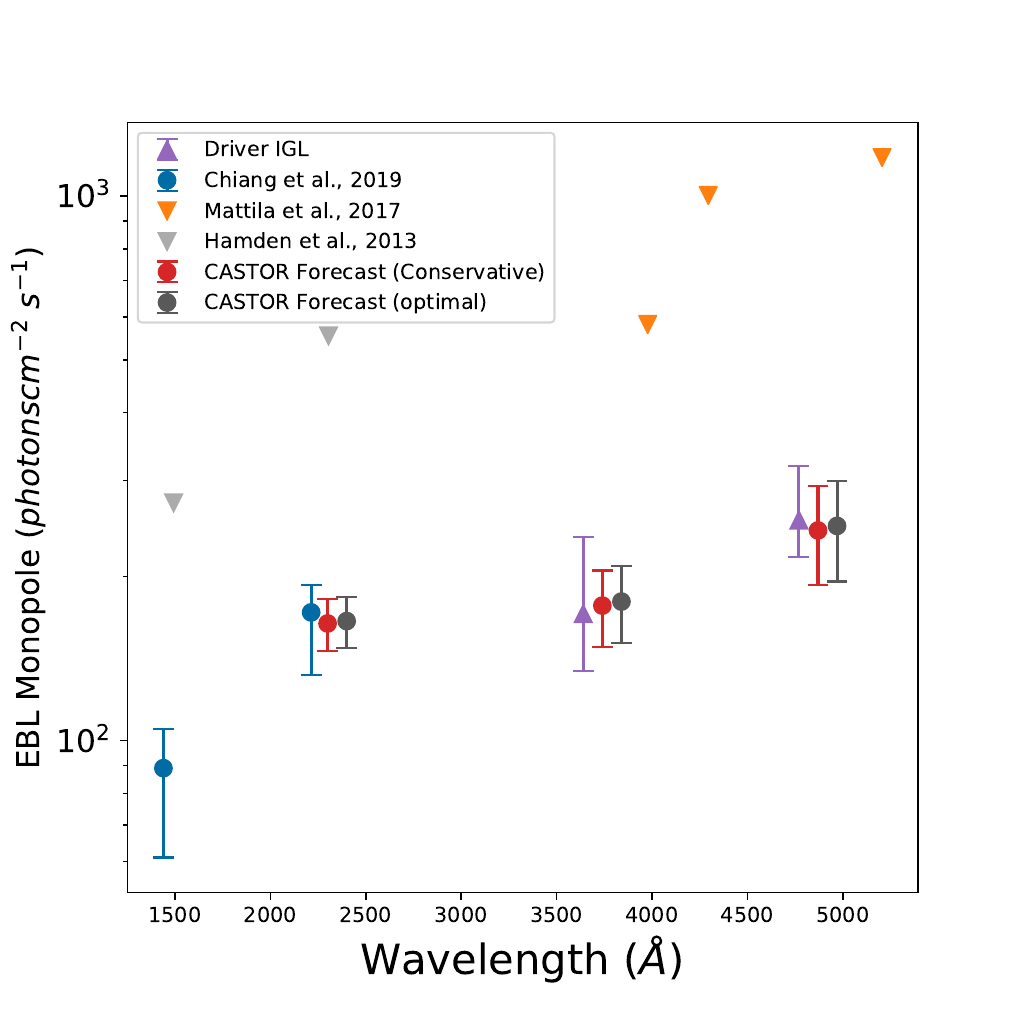}}
\caption{Comparison of our forecasted EBL monopole values in the uv, u and g {\small CASTOR} filters (red and dark grey bars) to the constraints on the intergalactic light (galaxies + AGN only) from \citet{Driver2016} (purple arrows), \citet{Mattila2017} using a dark cloud technique (orange arrows), the high galactic latitude measurements from \citet{Hamden2013} (light grey arrows) and the GALEX constraints (blue bars). To facilitate comparison of the uncertainties, we have introduced a 100 \AA\ offset in wavelength. As described in the text, we have also multiplied the \small{CASTOR} forecasted values by $\approx 1.4$, to better match the existing constraints given differences in the optical depth model. \label{fig:Literaturecomparison} }
\end{figure}

For our fiducial model, \cite{Chiang_2019} estimates values in the NUV of 172 photons cm$^{-2}$ s$^{-1}$ Hz, while \cite{Driver2016} gives lower limits of 171 and 254 photons cm$^{-2}$ s$^{-1}$ Hz in the u and g. The {\small CASTOR} uv filter nearly replicates the information present in the NUV filter on GALEX (see Figure~\ref{fig:filter_curves}), while the u and g filters extend these constraints into the blue end of the optical. Our forecast EBL monopoles in each filter and their associated $1-\sigma$ uncertainties determined from the posterior distributions of each model parameter are shown in Figure~\ref{fig:Literaturecomparison}. The measured quantity is the extragalactic light at the Earth; there is hence a degeneracy between the emitted extragalactic background light and intergalactic medium absorption. The EBL is thus measured up to a function of the mean optical depth $\tau_{\rm eff}$. Our simple analytic model for $\tau_{\rm eff}$ differs from the model of  \cite{Inoue2014}, which was used to derive the parameters of the fiducial model in \cite{Chiang_2019}, by $20-30\%$. To facilitate a comparison between our forecast and the existing results, we rescale our optical depth model so that the central forecast CASTOR monopole in the uv filter matches the measured value in the GALEX nuv filter. This corresponds to a factor of $\approx$ 1.4.


We compare our uncertainties on the EBL monopole to existing limits in Figure~\ref{fig:Literaturecomparison}. Our forecasted constraints are shown in red for the conservative model and dark grey for the optimal model. We also include the limits in the u and g derived from COSMOS, HST ERS, and HST UVUDF number and luminosity counts \citep{Driver2016}, including only extra-galactic contributions from discrete galaxies as purple lower limits. Also shown are constraints from dark cloud measurements \citep{Mattila2012} as orange upper limits, and observations at high galactic latitudes from \cite{Hamden2013} as light grey upper limits as they include contributions from both foregrounds and the extragalactic background.

Our error bars are derived under both optimal and conservative error models from the posterior fits to the spectral energy distribution. Both {\small GALEX} FUV and {\small CASTOR} uv, u, g have uncertainties about a factor of 3 smaller than the New Horizons measurements and comparable to the uncertainties from number count techniques, with the advantage of an unambiguous decomposition into extra-galactic and galactic components. \small{CASTOR's} redder filter set extends measurement of the EBL monopole as compared to \small{GALEX}. As the EBL monopole is already well constrained in \small{GALEX}, and it is not uniquely fixed by the emission redshift distribution, we do not find substantial additional constraining power. Similarly, as the EBL monopole is sensitive to the value of $\gamma_{\epsilon1500}$ that is not well constrained in either the conservative or optimal error models, there is little change in the uncertainty on the inferred monopole between the two CASTOR scenarios we consider.

Although these constraints are competitive with the current best constraints, we caution that our technique only measures the EBL monopole up to a degeneracy with the photon clustering bias that must be determined independently. This degeneracy can be broken if one has a priori knowledge of the emissivity distribution or, as in \cite{Chiang_2019}, a near-flat estimate of the slope of the continuum that produces an integral constraint on the bias normalization. Further, the EBL monopole is estimated only up to a factor dependent on the effective optical depth.

\section{SPHEREx and LUVOIR}
\label{sec:spherex}

{\small CASTOR} is able to constrain the EBL from $z= 0$ to $z \approx 3$ as compared to $z < 1$ for GALEX. Such a measurement is enabled by high redshift spectroscopic catalogs produced by ground based large scale structure surveys (DESI, eBOSS), and would be extended by complementary spectro-photometric observations with {\small SPHEREx} at higher redshift and {\small LUVOIR} at high spectral resolution.

In this section, we study an extension of our {\small CASTOR} models with {\small SPHEREx} using a simplified SED model. We then discuss the ability of {\small LUVOIR} to carry out a deep UV-optical intensity mapping experiment.

\subsection{SPHEREx Filters, Error Model, and Results}
\label{SPHEREx_results}

The {\small SPHEREx} instrument is a spectro-photometer based on a series of linear variable filters arranged such that the scan of the telescope across the sky exposes each independently. With accurate pointing knowledge, spectra for each point on the sky are reconstructed. Spectral resolution across the complete band-pass varies and is $R = 35 - 130$. 

{\small SPHEREx} Lyman-$\alpha$ intensity maps will be produced by observing from $0.75- 1.2 \mu$m with $R = 41$. We can model the spectro-photometer of {\small SPHEREx} as a limiting case of a broadband tomographic experiment where one defines a series of narrow-band filters that approximate the response of the spectro-photometer. The conservative instrument can be modelled by a series of 96 filters, however, only the first 19 constrain Ly$\alpha$ over this redshift range. We model these as a series of Gaussians with FWHM set by the spectral resolution.

With 19 effective narrowband photometric filters, inferring the parameters of the conservative rest frame emissivity model discussed in Section~\ref{sec:emissivity} becomes computationally expensive. To mitigate this, we consider only the terms governing emission of Ly$\alpha$ and the Ly continuum observed over this redshift range. We fix all other parameters to their fiducial values. In total, we constrain the bias evolution in frequency and redshift, $\gamma_{\nu}, \gamma_{z}$, as well as the Ly-continuum slope and evolution parameters $\alpha_{1216}$ and $C_{\alpha 1216}$ and Ly$\alpha$ equivalent width, evolved to its low redshift value, EW$_{Ly\alpha}^{z=1}$.

For {\small CASTOR}, existing spectroscopic tracer catalogs and robust estimates of the bias evolution allowed us to place limits on the signal to noise properties of our error budget at the $1-10\%$ level. For {\small SPHEREx}, few spectroscopic tracer objects are known at $z>5$ and future tracer catalog depths (from, e.g., Roman Space Telescope) are only known to within an order of magnitude \citep{spergel2015widefield}. The lack of reliable high redshift catalogs and constraints on the bias evolution makes a detailed estimate of the cross correlation errors depend on assumptions about the cosmological stellar mass and survey selection functions at high redshift. Rather than make assumptions about parameters which can vary over an order of magnitude, we instead consider fixed redshift independent fractional errors on estimates of the cross correlation and derive corresponding SED parameter constraints.

\begin{figure}

\includegraphics[width=0.485\textwidth]{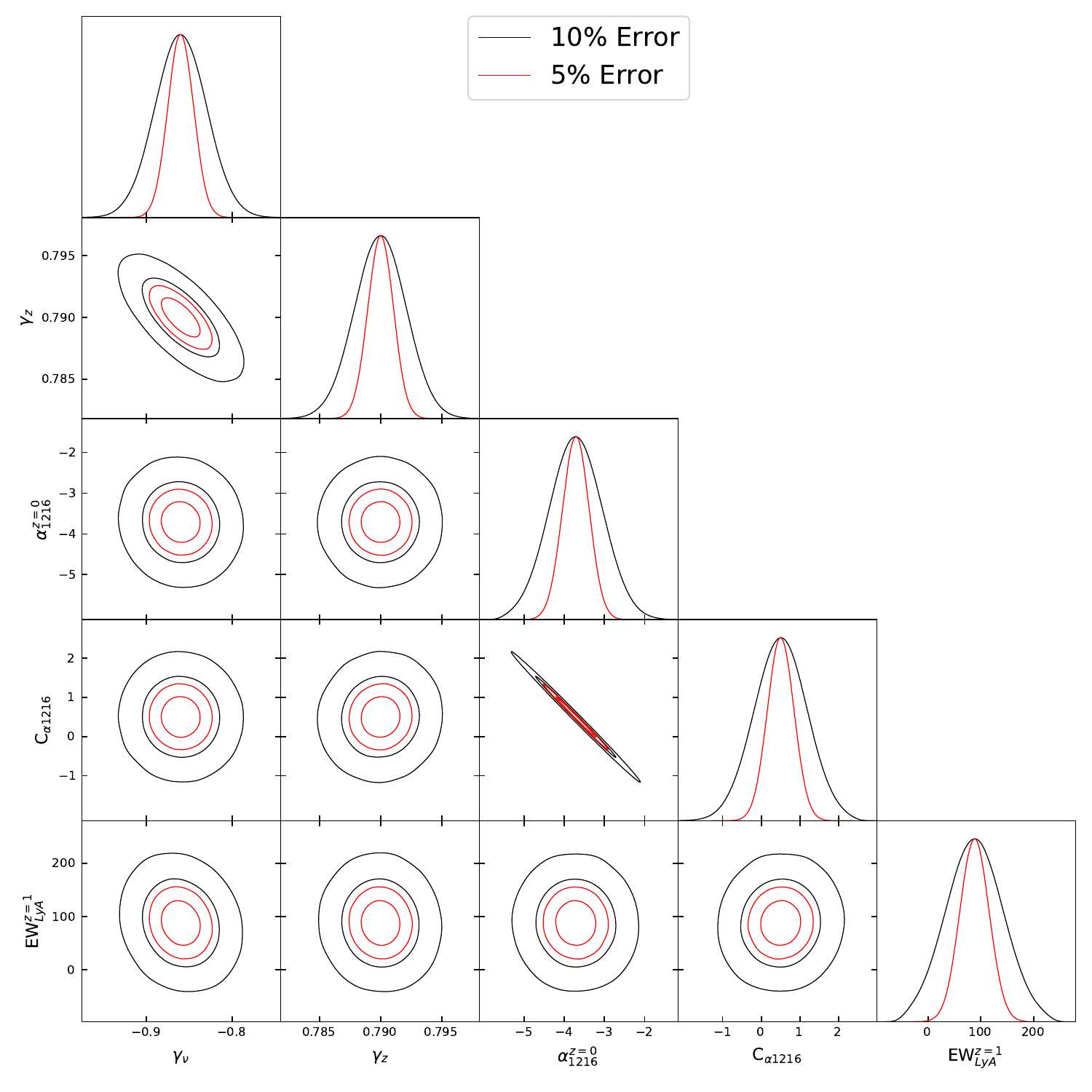}
\caption{\label{fig:SPHEREx_SED}Restricted parameter uncertainties on {\small SPHEREx} with fractional uncertainties of of $5\%$ and $10\%$.}
\end{figure}

We summarize results for {\small SPHEREx} in Figure~\ref{fig:SPHEREx_SED} for fractional errors of 5$\%$ and 10$\%$. The former approximates the average amplitude of the bootstrapped errors in {\small GALEX} while the former inflates these to approximate the peak observed noise amplitudes. Marginalized uncertainties are given in Table~\ref{tab:table3} as the $67\%$ interquartile range. Either error model produces $1\%$ level constraints on the frequency and redshift bias evolution, $10\%$ level constraints on the continuum slope, a $30-50\%$ constraint on $EW_{Ly\alpha}^{z=1}$ and constrains the continuum normalization to within an order of magnitude.

\begin{table}
  \begin{center}
      \caption{Posterior Interquartile Range on the parameters of a simplified high redshift SED model assuming fractional uncertainties of 5$\%$ and 10$\%$ for measurement of the bias weighted redshift distribution.\label{tab:table3}}
    \begin{tabular}{c c c c}
    
      \textbf{Parameter} & \textbf{10$\%$} & \textbf{5$\%$} & \textbf{Fiducial}\\
      \hline
      $\gamma_{\nu}$ & [-0.89, -0.83] & [-0.87, -0.85] & -0.86\\ 
      $\gamma_{z}$ & [0.788, 0.792] & [0.789, 0.791] & 0.79\\ 
      $\alpha_{1216}^{z=0}$ &  [-4.4, -3.02] & [-4.04, -3.38] & -3.71\\
      $C_{\alpha 1216}$ & [-0.22 ,1.22]&[0.16, 0.84] & 0.5\\
      $EW_{Ly\alpha}^{z=1}$ &[38.91 ,138.38] & [60.5, 115.52] & 88.02\\
    \end{tabular}
  \end{center}
\end{table}

The 30$\%$ constraint on the Ly$\alpha$ equivalent width provides a window into the population of Lyman $\alpha$ emitters (LAEs) at high redshift (\citealt{Ouchi2019} and references therein, \citealt{1967ApJ...147..868P}). Traditional techniques for studying this population rely on identifying individual sources with either narrowband photometry or slitless spectroscopy, both of which suffer from long exposure times on 4-8 meter class telescopes and limit the population of identified LAEs up to  $z\approx 8$ to $10^3-10^4$ \citep{Konno2016}. In contrast, our technique measures the population statistics of Ly$\alpha$ emitters at $z=5-9$ without identifying individual LAEs. Further, comparison of the bias parameters for Ly$\alpha$ sources and high redshift AGN can shed light on the relationship between LAEs and AGN through their environmental dependence \citep{Coil_2009,Sheth_1999}. The mass function $n(M)$ and luminosity function $\phi(L)$ are also constrained through their dependence on the total UV photon density $\rho_{\rm UV}$ and escape fraction $f_{\rm esc}$ that we measure through the evolution of the EBL monopole.

Previously, measurements of {\bf the rest-UV specific luminosity density and Ly$\alpha$ luminosity density, $\rho_{\rm UV}$ and $\rho_{{\rm Ly}\alpha}$,} relied on high resolution spectroscopy. However, we {\bf are able to} forecast constraints on these {\bf at the $10-30 \%$ level as determined from our posterior distribution for the SED model parameters, assuming our emissivity model can be extrapolated to the higher $z=5-9$ redshift range}. This particular redshift range is of interest, as they are expected to evolve rapidly at this epoch. Both $\rho_{\rm UV}$ and $\rho_{{\rm Ly}\alpha}$ can provide direct constraints on the timing and sources of reionization, so such a measurement overlapping with the expected reionization epoch expected to end around $z\approx 6-5.5$ \citep{fan06,kulkarni19,keating20,nasir20} would be of particular interest.

Population synthesis modeling of high redshift LAEs have been challenged by the lack of high signal to noise continuum detection in individual spectra \citep{Lai2008,Bruzual2003}. In contrast to stacking techniques that are challenged by the presence of strong nebular lines, our technique is sensitive only to broadband noise features that are produced systematically across the population of LAEs. The joint analysis of a broadband tomographic measurement of the LAE population and high spectral resolution studies of samples of LAEs would benefit from differences in the underlying systematic uncertainties.

\subsection{Measuring the UV-Optical Background SED with LUVOIR} 

In contrast to {\small SPHEREx} and {\small CASTOR}, the LUMOS instrument on {\small LUVOIR} will enable studies of a small number of sources with extremely high spectral resolution \citep{LUVOIRfinal}. Studies of the IGM and CGM will primarily use background QSOs to study the diffuse sky in absorption from $z=1-2$. Systematic and statistical errors in measurements of absorption constraints on the UV background are summarized in \citep{Becker2013}, and include uncertainties in the effective optical depth, temperature-overdensity relation, and Jean's smoothing, which affect small scales and require detailed modeling or simulations to estimate. In contrast, our error budget is dominated by large scale effects that evolve with redshift and can be estimated with linear theory.

In addition to the spectroscopic instrument, {\small LUVOIR} also has a UV-optical High Definition Imager (HDI) with wavelength coverage from 0.2-2.5 $\mu$m. Although a galaxy counting experiment with {\small LUVOIR} would have improved depth compared to HST measurements, it would be limited to the component of the UV background which arises directly from discrete components. A diffuse sky measurement similar to {\small GALEX} and the one we envision with {\small CASTOR} could provide a powerful complement to measurements of discrete sources. Broadband tomography will yield competitive constraints if a large enough area of the sky can be observed such that there is sufficient deep spectroscopic tracer catalog overlap and minimal uncertainties due to cosmic or sample variance. Compared to a large focal plane survey mission, HDI has a field of view of $0.2' \times 0.3'$, about a factor of $150$ smaller than the 0.25 deg$^2$ field of view for the {\small CASTOR} imager.

{\small LUVOIR}-A is envisioned to have a 15 m primary, while {\small LUVOIR}-B would have a more modest 8 m primary, corresponding to factors of 225 and 64 in light gathering power, respectively. Assuming a constant limiting magnitude equivalent for both, an intensity mapping experiment with {\small LUVOIR} would then survey an equivalent area with a similar overall exposure time. A $1$ month intensity mapping survey with {\small LUVOIR} would scan a map of $\approx 100 \mathrm{deg}^2$. Such a survey would likely be cosmic variance limited at the $1\%$ level assuming the scalings in \citep{Moster_2011}. Since only knowledge, but not control, of telescope pointing is required, it is interesting to note that a 6 month survey with the Hubble Space Telescope following gyroscope failure could achieve a similar level of constraining power.

\section{Conclusions} 
\label{sec:conclusions}

We have considered the ability of future survey instruments, {\small CASTOR} and {\small SPHEREx}, and optionally {\small LUVOIR} or HST, to extend the constraints from {\small GALEX} on the extragalactic component of the optical and UV background light at redshifts $z=0-3$ and $z=5-9$. The low redshift regime constrains properties of the UV background and the high redshift regime constrains the timing and sources of reionization. For {\small CASTOR}, we have modelled measurement uncertainties with a combination of shot noise from galaxy cross-correlation tracers, photometric errors, and fluctuations in the bias evolution. We consider two error models, a limiting optimal model achievable with future spectroscopic catalogs where shot noise from the tracer catalog is subdominant to photometric errors, and a conservative model intended to bound upper limits on each effect we consider. For {\small SPHEREx}, we instead considered fixed total error budgets of 5$\%$ and 10$\%$. We derive posterior distributions on the model parameters for each model and experiment.

For {\small CASTOR}, we find a factor of $2-3$ improvement in the geometric mean of the relative errors in parameters of our spectral energy distribution model under conservative and optimistic error models respectively. These constraints are determined from the application of clustering redshift estimation to a future all sky broadband intensity mapping experiment. From the posterior SED fits, we estimated monopole uncertainties for the uv, u and g filters, finding that these constraints are competitive under both error models. {\small SPHEREx} would constrain Ly$\alpha$ emission at the $10-30\%$ level from $z=5-9$ and shed light on the population of Ly$\alpha$ emitters at high redshift.

An observed frame UV broadband tomographic measurement with {\small CASTOR} intensity maps would represent a significant improvement on current experiments targeting these wavelengths. {\small SPHEREx}, by contrast, would constrain the population of Lyman-$\alpha$ emitters at high redshift with observed frame infrared measurements. 
Intriguingly, {\small LUVOIR}'s large mirror size compensates for its small field of view and would allow it to place tight limits on the UV-optical SED with a modest investment in observing time. Similarly, since accurate pointing control is not necessary, a larger investment of HST time in a post-gyroscopic failure mode offers a promising extension to this storied mission's history as a photometric intensity mapping experiment.

{\small CASTOR} and {\small SPHEREx} would yield an improved picture of the low surface brightness universe and total photon budget in two windows, from $z=1-3$ and $z=5-9$. Together, we expect knowledge of the SED at the few percent level, representing a factor of $2-10$ over the current state of the art.

\section*{Acknowledgments}

PUS and SB were supported by NSF grant AST-1817256. We thank Peter Capak, Daniel Masters, Brian Siana, and Anson D'Aloisio for helpful conversations. Ming-Feng Ho and Patrick C\^ot\'e provided helpful comments on an earlier draft of this manuscript. We thank the anonymous referee for their helpful comments and thorough reading of the text.

\section*{Data Availability}

All data is available publicly at \url{https://github.com/bscot/Broadband_tomography_with_CASTOR_and_SPEHREx}

\bibliographystyle{mnras}
\bibliography{castor_bibtex}{}

\bsp	
\label{lastpage}

\end{document}